\documentclass[creativecommons]{eptcs}

\usepackage[utf8]{inputenc}

\usepackage{algorithm}
\usepackage[noend]{algpseudocode}
\usepackage{amsmath}
\usepackage{amssymb}
\usepackage{amsthm}
\usepackage{booktabs}
\usepackage{enumitem}
\usepackage{mathpartir}
\usepackage{microtype}
\usepackage{relsize}
\usepackage{stmaryrd}
\usepackage{thm-restate}

\usepackage{tikz}
\usetikzlibrary{positioning,calc}

\renewcommand{\mid}{\,\,|\,\,}

\newcounter{theoremcnt}[section]

\theoremstyle{definition}

\newtheorem{example}[theoremcnt]{Example}
\newtheorem{definition}[example]{Definition}
\newtheorem{lemma}[definition]{Lemma}

\newcommand{\eg}{\emph{e.g.}}
\newcommand{\ie}{\emph{i.e.}}

\newcommand{\lscott}{\ensuremath{[\![}}
\newcommand{\rscott}{\ensuremath{]\!]}}

\newcommand{\multiset}[1]{\Lbag #1 \Rbag}
\newcommand{\set}[1]{{\{ #1 \}}}
\newcommand{\sem}[1]{\mathbf{#1}}
\DeclareMathOperator*{\bigplus}{\mathlarger{\mathlarger{\mathlarger{+}}}}

\newcommand{\booleans}{\mathbb{B}}
\newcommand{\naturalnumbers}{\mathbb{N}}

\newcommand{\subst}{\sigma}

\newcommand{\interpret}[1]{\ensuremath{\lscott #1\mkern1mu\rscott}}
\newcommand{\multiact}{\alpha}
\newcommand{\nodata}[1]{\underline{#1}}
\newcommand{\sequential}{\mathbin{.}}

\newcommand{\allow}{\nabla}
\newcommand{\communication}{\Gamma}
\newcommand{\comms}{\mathsf{Comm}}
\newcommand{\hide}{\tau}

\newcommand{\actcomm}[1]{\gamma_{#1}}
\newcommand{\acthide}[1]{\theta_{#1}}
\newcommand{\freevars}{\textsf{FV}}

\newcommand{\lts}{\mathcal{L}}
\newcommand{\states}{\mathit{S}}
\newcommand{\events}{\Lambda}

\newcommand{\actions}{\mathit{Act}}
\newcommand{\semactions}{\Omega}
\newcommand{\action}{\omega}

\newcommand{\transitions}{\mathbin{\rightarrow}}
\newcommand{\transition}[1]{\xrightarrow{#1}}

\newcommand{\simpleprocess}{\mathsf{S}}
\newcommand{\linearprocess}{\mathsf{P}}
\newcommand{\names}{\mathit{PN}}
\newcommand{\initval}{\iota}

\newcommand{\actsync}{\mathsf{sync}}
\newcommand{\acttag}{\mathsf{tag}}
\newcommand{\Iind}{K}
\newcommand{\Isub}{J}

\newcommand{\REQ}[1]{\ifx1#1\textsc{SYN}\else\ifx2#1\textsc{IND}\else\ifx3#1\textsc{ORI}\else\ifx4#1\textsc{COM}\fi\fi\fi\fi}

\newcommand{\bisim}{\mathbin{\leftrightarroweq}}

\newcommand{\related}{\mathbin{R}}

\newcommand{\boolsort}{\mathit{Bool}}
\newcommand{\natsort}{\mathit{Nat}}

\newcommand{\false}{\textsf{false}}
\newcommand{\true}{\textsf{true}}

\newcommand{\actcount}{\textsf{count}}
\newcommand{\acttoggle}{\textsf{toggle}}
\newcommand{\actdrill}{\textsf{drill}}

\newcommand{\procmachine}{\text{Machine}}
\newcommand{\procdrill}{\text{Drill}}

\newcommand{\subvector}[2]{{#1}_{|#2}}
\newcommand{\flatten}[1]{\mathsf{Vars}(#1)}
\renewcommand{\vector}[1]{\langle #1 \rangle}

\newcommand{\chatbox}{\textsf{Chatbox}}
\newcommand{\hesselink}{\textsf{Register}}
\newcommand{\WMS}{\textsf{WMS}}
\newcommand{\abp}{\textsf{ABP}}
\newcommand{\connectfour}{\textsf{Connect Four}}

\title{Decomposing Monolithic Processes in a Process Algebra with Multi-actions}
\def\titlerunning{Decomposing Monolithic Processes in a Process Algebra with Multi-actions}
\def\authorrunning{M. Laveaux and T.A.C.~Willemse}

\author{Maurice Laveaux \qquad \qquad\qquad Tim A.C.~Willemse
\institute{Eindhoven University of Technology,\\  Eindhoven, The Netherlands}
\email{m.laveaux@tue.nl \quad  t.a.c.willemse@tue.nl}
}

\begin{document}
\maketitle

\begin{abstract}
A monolithic process is a single recursive equation with data parameters, which only uses non-determinism, action prefixing, and recursion.
We present a technique that decomposes such a monolithic process into multiple processes where each process defines behaviour for a subset of the parameters of the monolithic process.
For this decomposition we can show that a composition of these processes is strongly bisimilar to the monolithic process under a suitable synchronisation context.
Minimising the resulting processes before determining their composition can be used to derive a state space that is smaller than the one obtained by a monolithic exploration.
We apply the decomposition technique to several specifications to show that this works in practice.
Finally, we prove that state invariants can be used to further improve the effectiveness of this decomposition technique.
\end{abstract}

\section{Introduction}

The mCRL2 language~\cite{GrooteM2014} is a process algebra that can be used to specify the behaviour of communicating processes with data parameters.
It has the usual ACP-style operators for modelling non-deterministic choice, sequential composition, parallel composition and recursion.
A powerful yet somewhat unconventional language construct of mCRL2 is the \emph{multi-action}, which allows for specifying that atomic actions can happen simultaneously.

Specifications written in mCRL2 can be analysed using the mCRL2 toolset~\cite{BunteGKLNVWWW19}.
The corresponding mCRL2 toolset~\cite{BunteGKLNVWWW19} translates a process specification to an equivalent monolithic recursive process, replacing all interleaving parallelism by non-determinism, action prefixing and recursion.
Translating a complicated process specification into a simpler normal form, in this case the monolithic process, has several advantages.
First of all, the design and implementation of state space exploration algorithms can be greatly simplified.
Furthermore, the design of effective static analysis techniques on the global behaviour of the specification is also easier.
One example is a static analysis to detect live variables as presented in~\cite{PolT09}.
However, the static analysis techniques available at the moment are not always strong enough to mitigate the state space explosion problem for this monolithic process even though its state space can often be minimised with respect to some equivalence relation after state space exploration.

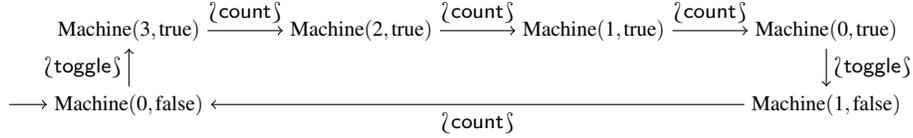
\begin{figure}[t]
	\centering
	\begin{tikzpicture}[->]
 		\scriptsize
	  \node (s0) {$\procmachine(0, \text{false})$};
	  \node (init) [left=5mm of s0] {};
	  \draw (init) edge (s0);
	      
	  \node (s1) [above=.5cm of s0] {$\procmachine(3, \text{true})$};
	  \node (s2) [right=of s1] {$\procmachine(2, \text{true})$};
	  \node (s3) [right=of s2] {$\procmachine(1, \text{true})$};
	  \node (s4) [right=of s3] {$\procmachine(0, \text{true})$};
	  \node (s5) [below=.5cm of s4] {$\procmachine(1, \text{false})$};

	  \draw (s0) edge node[left] {$\multiset{\acttoggle}$} (s1);	  
	  \draw (s1) edge node[above] {$\multiset{\actcount}$} (s2);
	  \draw (s2) edge node[above] {$\multiset{\actcount}$} (s3);
	  \draw (s3) edge node[above] {$\multiset{\actcount}$} (s4);
	  \draw (s4) edge node[right] {$\multiset{\acttoggle}$} (s5);
	  \draw (s5) edge node[below] {$\multiset{\actcount}$} (s0);  
	\end{tikzpicture}	
  \caption{Behaviour of a machine.}\label{figure:delayedmachine}
\end{figure}

\begin{figure}[t]
\begin{center}
\begin{tikzpicture}[->]
	\scriptsize
 \node (s0) {$\procmachine_V(0)$};
 \node (s0init) [left= 0.5cm of s0] {};
 \node (s1) [above right= 0.5cm of s0] {$\procmachine_V(3)$};
 \node (s2) [below right= 0.5cm of s1] {$\procmachine_V(2)$};
 \node (s3) [below right= 0.5cm of s0] {$\procmachine_V(1)$};

 \draw (s0init) edge (s0);
 \draw (s0) edge[bend left=20] node[above left] {$\multiset{\actsync^1_V(\textbf{\false})}$} (s1);	
 \draw (s0) edge[bend left=20] node[above right,pos=-0.1] {$\multiset{\actsync^1_V(\textbf{\true})}$} (s3);	  
 \draw (s1) edge[bend left=20] node[above right] {$\multiset{\actcount,\acttag}$} (s2);
 \draw (s2) edge[bend left=20] node[below right] {$\multiset{\actcount,\acttag}$} (s3);  
 \draw (s3) edge[bend left=20] node[below left] {$\multiset{\actcount,\acttag}$} (s0);
  
 \node (t0) [right=3cm of s2] {$\procmachine_W(\false)$};
 \node (t0init) [left= 0.5cm of t0] {};
 \node (t1) [right= of t0] {$\procmachine_W(\true)$};
 	
 \draw (t0init) edge (t0);
 \draw (t0) edge[bend left] node[above] {$\multiset{\acttoggle, \actsync^1_W(\textbf{\false})}$} (t1);	  
 \draw (t1) edge[bend left] node[below] {$\multiset{\acttoggle, \actsync^1_W(\textbf{\true})}$} (t0);
\end{tikzpicture}
\end{center}
\caption{Behaviour of the decomposition processes.}\label{figure:decomposition}
\end{figure}
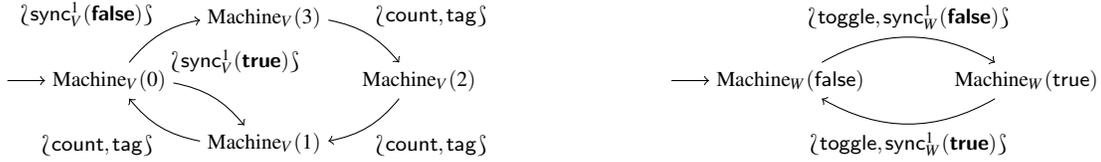

In this paper, we define a decomposition technique (which we refer to as a \emph{cleave}) of a monolithic process. Our technique takes as input such a process and a partitioning of its data parameters, and it produces two new processes.
To illustrate the idea, consider a machine that alternates between two modes, where switching modes has a certain delay.
The behaviour of this machine is modelled by the labelled transition system in Figure~\ref{figure:delayedmachine}.
Assume that this machine is described by a single recursive mCRL2 process with two parameters: a natural number representing the counter and a Boolean for representing the mode of the machine.
Using the partition that `splits' these two parameters, our technique will decompose this machine into two recursive processes (\emph{components}) with their respective behaviour shown in Figure~\ref{figure:decomposition}.
Observe that indeed the states of a component rely on only one of the two parameters.
Furthermore, note that the transition systems of both components include $\actsync$ actions that do not occur in the transition system of the original machine.
These are generated by our technique and are needed to model the \emph{interface} between the two components such that
under a suitable synchronisation context the parallel composition of these components \emph{is} equivalent (\emph{strongly bisimilar}) to the original monolithic process.

Decomposing a monolithic process may help to partly sidestep the \emph{state space explosion} that is due to the interleaving of parallel processes that is encoded in the monolithic process.
This follows from the observation that the state spaces of the components resulting from a decomposition can be (much) smaller than the state space of the monolithic process; these may therefore be easier to obtain.
By first minimising the state spaces of these components with respect to bisimilarity \emph{before} computing their composition, also the state space of the latter remains (much) smaller than that of the monolithic process.
Since strong bisimilarity is a congruence for all operators of mCRL2, the resulting state space is still strongly bisimilar to that of the monolithic process, meaning that no information is lost.

Theoretically, the main challenge in defining a decomposition technique is to ensure that it results in components that, when combined appropriately, behave indistinguishably from the monolithic process from which they were derived.
This is the problem of finding a \emph{valid decomposition}.
We illustrate that there may be multiple valid decompositions of a monolithic process.
The main practical challenge is therefore to identify a universally applicable decomposition technique that yields valid decompositions, and which is capable of sidestepping the state space explosion problem.
Summarising, the contributions of our work are as follows:
\begin{itemize}
	\item we formalise the notion of a decomposition and the notion of validity of a decomposition, 
  
  \item we present a generally applicable decomposition technique and provide sufficient conditions for this decomposition to be valid,
	
	\item we show that \emph{state invariants}~\cite{GrooteM92} can be used to obtain even smaller state spaces by restricting the interfaces of the components resulting from the decomposition,
	
	\item we confirm the practical applicability of our techniques on several cases.
\end{itemize}

\paragraph{Related Work.}
Several different techniques are related to this type of decomposition.
Most notably, the work on decomposing Petri nets into a set of automata~\cite{BouvierGL20} also aims to speed up state space exploration by means of decomposition.
The work on functional decomposition~\cite{BrinksmaLB93} describes a technique to decompose a specification based on a partitioning of the action labels instead of a partitioning of the data parameters.
In~\cite{JongmansCP16} it was shown how this type of decomposition can be achieved for mCRL2 processes.
Furthermore, a decomposition technique was used in~\cite{GrooteM92} to improve the efficiency of equivalence checking.
However, that work considers processes that are already in a parallel composition and further decomposes them based on the actions that occur in each component.

Decompositional minimisation is also related to \emph{compositional minimisation}, in which the objective is to replace the state space of each component in a (given) parallel composition by an equivalent, smaller state space, while preserving the behaviour of the original specification~\cite{TaiK93:incremental,TaiK93:hierarchy}.
A problem that is common to compositional minimisation and decompositional minimisation is that the size of the state spaces belonging to individual components summed together might exceed the size of the original state space~\cite{GaravelLM18}.
One way to (partly) avoid this is by specifying \emph{interface constraints} (also known as \emph{environmental constraints} or \emph{context constraints}), see~\cite{GrafSL96,CheungK96}.
Note that the state invariants in our work serve a similar purpose, but the mechanism is different since interface constraints are action-based whereas invariants are state-based.
Another possibility is to find a more suitable order in which components are explored and minimised, since the order heavily influences the size of the intermediate state spaces.
Heuristics for determining this order can be very effective in practice~\cite{CrouzenL11}; such heuristics are also relevant for the application of our decomposition technique.

One advantage of the decomposition technique over compositional minimisation is that our interfaces can be derived from the conditions present in the monolithic process.
These interfaces can also be further strengthened with state invariants.
Secondly, the components resulting from the decomposition are not limited to the user-defined processes present in the specification.
Our decomposition technique is thus more flexible, and may yield more optimal compositions.
Indeed, the case studies on which we report support both observations.

\paragraph{Outline.} 
In Section~\ref{section:preliminaries} the syntax and semantics of the considered process algebra are defined.
The decomposition problem is defined in Section~\ref{section:decomposition} and the cleave technique is presented in Section~\ref{section:solution}.
In Section~\ref{section:state_invariant} the cleave technique is improved with state invariants.
In Section~\ref{section:implementation} the implementation is described shortly and a case study is presented in Section~\ref{section:casestudy} to illustrate the effectiveness of the decomposition technique in practice.
Finally, a conclusion and future work is presented in Section~\ref{section:conclusion}.

\section{Preliminaries}\label{section:preliminaries}

We assume the existence of an abstract data theory that describes data sorts.
Each sort $D$ has an associated non-empty semantic domain denoted by $\mathbb{D}$.
The existence of sorts $\boolsort$ and $\natsort$ with their associated Boolean ($\booleans$) and natural number ($\naturalnumbers$) semantic domains respectively, with standard operators is assumed.
Furthermore, we assume the existence of an infinite set of \emph{sorted variables}.
We use $e : D$ to indicate that $e$ is an expression (or variable) of sort $D$.
The set of free variables of an expression $e$ is denoted $\freevars(e)$, and
a variable that is not free is called \emph{bound}.
An expression $e$ is \emph{closed} iff $\freevars(e) = \emptyset$.
A substitution $\subst$ is a total function from variables to closed data expressions of their corresponding sort.
We use $\subst(e)$ to denote the syntactic replacement of variables in expression $e$ by their substituted expression.

An \emph{interpretation} function, denoted by $\interpret{\ldots}$, maps syntactic objects to values within their corresponding semantic domain.
We assume that $\interpret{e}$ for closed expressions $e$ is already defined.
Semantic objects are typeset in \emph{boldface} to differentiate them from syntax, \eg, the semantics of expression $1 + 1$ is $\textbf{2}$.
We denote \emph{data equivalence} by $e \approx f$, which is $\true$ iff $\interpret{e} = \interpret{f}$;
for other operators we use the same symbol in both syntactic and semantic domains.
We adopt the usual principle of substitutivity; \ie, for all variables $x$, expressions $e$ and closed expressions $g$ and $h$ it holds that if $g \approx h$ then $[x \gets g](e) \approx [x \gets h](e)$.

We denote a \emph{vector} of length $n+1$ by $\vec{d} = \vector{d_0, \ldots, d_n}$.
Two vectors are equivalent, denoted by $\vector{d_0, \ldots, d_n} \approx \vector{e_0, \ldots, e_n}$, iff their elements are \emph{pairwise equivalent}, \ie, $d_i \approx e_i$ for all $0 \leq i \leq n$.
Given a vector $\vector{d_0, \ldots, d_n}$ and a subset $I \subseteq \naturalnumbers$, we define the \emph{projection}, denoted by $\subvector{\vector{d_0, \ldots, d_n}}{I}$, as the vector $\vector{d_{i_0}, \ldots, d_{i_l}}$ for the largest $l \in \naturalnumbers$ such that $i_0 < i_1 < \ldots < i_l \leq n$ and $i_k \in I$ for $0 \leq k \leq l$.
We write $\vec{d} : \vec{D}$ for a vector of $n + 1$ variables $d_0 : D_0, \ldots, d_n : D_n$ and denote the projection for a subset of indices $I \subseteq \naturalnumbers$ by $\subvector{\vec{d}}{I} : \subvector{\vec{D}}{I}$.
Finally, we define $\flatten{\vec{d}} = \{d_0, \ldots, d_n\}$.

A \emph{multi-set} over a set $A$ is a total function $m : A \rightarrow \naturalnumbers$;
we refer to $m(a)$ as the \emph{multiplicity} of $a$ and we write $\multiset{\ldots}$ for a multi-set where the multiplicity of each element is either written next to it or omitted when it is one. 
For instance, $\multiset{a : 2, b}$ has elements $a$ and $b$ with multiplicity two and one respectively, and all other elements have multiplicity zero.
For multi-sets $m, m' : A \rightarrow \naturalnumbers$, we write 
$m \subseteq m'$ iff $m(a) \leq m'(a)$ for all $a \in A$. Multi-sets
$m+m'$ and $m-m'$ are defined pointwise: $(m + m')(a) = m(a) + m'(a)$ and  $(m - m')(a) = \text{max}(m(a) - m'(a), 0)$ for all $a \in A$.

\subsection{Labelled Transition Systems}

Let $\events$ be the set of (sorted) action \emph{labels}.
We use $D_a$ to indicate the sort of action label $a \in \events$.
The set of all multi-sets over $\{ a(\sem{e}) \mid a \in\events, \sem{e} \in \mathbb{D}_a\}$ is denoted $\semactions$.
Note that $\mathbb{D}_a$ is the semantic domain of $D_a$.
In examples we typically omit the expression and parentheses whenever $D_a$ consists of a single element.

\begin{definition}
  A labelled transition system with multi-actions, abbreviated LTS, is a tuple $\lts = (\states, s_0, \actions, \transitions)$ where $\states$ is a set of states; $s_0 \in S$ is an initial state; $\actions \subseteq \semactions$ and $\transitions\,\subseteq \states \times \actions \times \states$ is a labelled transition relation.
\end{definition}

We typically use $\action$ to denote an element of $\actions$ and we write $s \transition{\action} t$ whenever $(s, \action, t) \in \transitions$.
As usual, a finite LTS can be depicted as an edge-labelled directed graph, where vertices represent states, the labelled edges represent the transitions, and a dangling arrow indicates the initial state.
The left graph of Figure~\ref{figure:decomposition} depicts an LTS with four states and five transitions, which are
labelled with multi-actions $\multiset{\actcount,\acttag}, \multiset{\actsync_V^1(\textbf{\true})}$ and $\multiset{\actsync_V^1(\textbf{false})}$.

We recall the well-known strong bisimulation equivalence relation on LTSs~\cite{Milner83}.

\begin{definition}
  Let $\lts_i = (\states_i, s_i, \actions_i, \transitions_i)$ for $i \in \{1,2\}$ be two LTSs.
  A binary relation $R \subseteq \states_1 \times \states_2$ is a \emph{(strong) bisimulation relation} iff for all $s \related t$:
  \begin{itemize}
 		\item if $s \transition{\action}_1 s'$ then there is a state $t' \in \states_2$ such that $t \transition{\action}_2 t'$ and $s' \related t'$, and
 		
 		\item if $t \transition{\action}_2 t'$ then there is a state $s' \in \states_1$ such that $s \transition{\action}_1 s'$ and $s' \related t'$.
  \end{itemize}

  \noindent
  States $s$ and $t$ are \emph{bisimilar}, denoted $s \bisim t$, iff $s \related t$ for a bisimulation relation $R$.
  We write $\lts_1 \bisim \lts_2$ iff $s_1 \bisim s_2$ and say $\lts_1$ and $\lts_2$ are bisimilar.  
\end{definition}

\subsection{Linear Process Equations}

We draw inspiration from the process algebra mCRL2~\cite{GrooteM2014}, which contains multi-actions, to describe the elements of an LTS;
similar concepts and constructs may appear in other shapes elsewhere.
\begin{definition}
  \emph{Multi-actions} are defined as follows:
  \begin{equation*}
    \multiact ~::=~ \tau ~\mid~ a(e) ~\mid~ \multiact | \multiact
  \end{equation*}
  Constant $\tau$ represents the empty multi-action and $a \in \events$ is an action label with an expression $e$ of sort $D_a$.
 	The semantics of a multi-action $\multiact$ for any substitution $\sigma$, denoted by $\interpret{\multiact}_\sigma$, is an element of $\semactions$ and defined inductively as follows: $\interpret{\tau}_\sigma = \multiset{}$, $\interpret{a(e)}_\sigma = \multiset{a(\interpret{\sigma(e)})}$ and $\interpret{\alpha | \beta}_\sigma = \interpret{\alpha}_\sigma + \interpret{\beta}_\sigma$.
  If $\multiact$ is a closed expression then the substitution can be omitted.
\end{definition}

The states and transitions of an LTS are described by means of a monolithic process called a \emph{linear process equation}, which consists of a number of \emph{condition-action-effect} statements, referred to as \emph{summands}.
Each summand symbolically represents a partial transition relation between the current and the next state for a multi-set of action labels.
Let $\names$ be a set of process \emph{names}.

\begin{definition}
  A \emph{linear process equation} (LPE) is an equation of the form:
  \[P(d : D) = \sum_{e_0 : E_0} c_0 \rightarrow \multiact_0 \sequential P(g_0) 
  + ~\ldots~
  + \sum_{e_n : E_n} c_n \rightarrow \multiact_n \sequential P(g_n)
  \]
  Where $P \in \names$ is the process \emph{name}, $d$ is the process parameter, and each:
  \begin{itemize}  
    \item $E_i$ is a sort ranged over by \emph{sum} variable $e_i$ (where $e_i \neq d$),

    \item $c_i$ is the \emph{enabling condition}, a boolean expression so that $\freevars(c_i) \subseteq \set{d, e_i}$, 

    \item $\multiact_i$ is a multi-action $\tau$ or $a^1_i(f^1_i) | \ldots | a^{n_i}_i(f^{n_i}_i)$ such that each $a^k_i \in \events$ and $f^k_i$ is an expression of sort $D_{a^k_i}$ such that $\freevars(f^k_i) \subseteq \set{d, e_i}$, 

    \item $g_i$ is an \emph{update} expression of sort $D$, satisfying $\freevars(g_i) \subseteq \set{d, e_i}$.
  \end{itemize}
\end{definition}

The $+$-operator denotes a non-deterministic choice among the summands of the LPE; the $\sum$-operator describes a non-deterministic choice among the possible values of the associated sum variable bound by the $\sum$-operator.
We omit the $\sum$-operator when the sum variable does not occur freely within the condition, action and update expressions.
We use $\bigplus_{i \in I}$ for a finite set of \emph{indices} $I \subseteq \naturalnumbers$ as a shorthand for a number of summands.

We often consider LPEs where the parameter sort $D$ represents a \emph{vector}; in that case we write $d_0 : D_0, \ldots, d_n : D_n$ to indicate that there are $n + 1$ parameters where each $d_i$ has sort $D_i$.
Similarly, we also generalise the action sorts and the sum operator in LPEs, where we permit ourselves to write $a(f_0, \ldots, f_k)$ and $\sum_{e_0:E_0, \ldots, e_l:E_l}$, respectively.\smallskip

The \emph{operational semantics} of an LPE are defined by a mapping to an LTS.
Let $\linearprocess$ be the set 
of symbols $P(\initval)$ such that $P(d : D) = \phi_P$, for any $P \in \names$, is an LPE and $\initval$ is a closed expression of sort $D$.

\begin{definition}\label{def:lpesemantics}
  Let $P(d : D) = \bigplus_{i \in I} \sum_{e_i : E_i} c_i \rightarrow \multiact_i \sequential P(g_i)$ be an LPE and let $\initval : D$ be a closed expression. The semantics of $P(\initval)$, denoted by $\interpret{P(\initval)}$, is the LTS $(\linearprocess, P(\initval), \semactions, \transitions)$ where $\transitions$ is defined as follows:
  for all indices $i \in I$, closed expressions $\initval' : D$ and substitutions $\subst$ such that $\subst(d) = \initval'$ there is a transition $P(\initval') \xrightarrow{\interpret{\subst(\multiact_i)}} P(\subst(g_i))$ iff $\interpret{\subst(c_i)} = \textbf{\true}$.
\end{definition}

For a given LPE, we refer to the reachable part of the LTS, induced by the LPE, as the \emph{state space}.
Note that in the interpretation of an LPE a syntactic substitution is applied to the update expressions to define the reached state.
This means that different closed syntactic expressions which correspond to the same semantic object, \eg, $1 + 1$ and $2$ for our assumed sort $\natsort$, result in different states.
As stated by the lemma below, such states are always bisimilar.

\begin{lemma}\label{lemma:bisimulation}
	Given an LPE $P(d : D) = \phi_P$.
	For all closed expressions $e, e' : D$ such that $\interpret{e \approx e'} = \textbf{\true}$ we have $\interpret{P(e)} \bisim \interpret{P(e')}$.	
\end{lemma}

For any given state space we can therefore consider a \emph{representative} state space where for each state a unique closed expression is chosen that is data equivalent.
In examples we always consider the representative state space.

\begin{example}\label{example:delayedmachine}
Consider the following LPE, modelling a machine that alternates between two modes.
The event that signals a switch between these two modes is modelled by action $\acttoggle$;
switching between modes happens after a number of clock cycles and is dependent on the mode the machine is running (3 cycles for one mode, 1 cycle for the other).
The machine keeps track of its mode using a Boolean parameter $s$, and a parameter $n$ which keeps track of the number of cycles left before switching modes.
\begin{align*}
	\procmachine(n : \natsort, s : \boolsort) 
   & = (n > 0) \rightarrow \actcount \sequential \procmachine(n - 1, s) \\
   & + (n \approx 0) \rightarrow \acttoggle \sequential \procmachine(\textsf{if}(\neg s, 3, 1), \neg s)
\end{align*}
Note that the expression $\textsf{if}(\neg s, 3, 1)$ models the reset of the clock cycle count upon switching modes.
A representative state space of the machine that is initially off, defined by $\interpret{\procmachine(0, \false)}$, is shown in Figure~\ref{figure:delayedmachine}.
\end{example}

\subsection{A Process Algebra of Communicating Linear Process Equations}
\label{sec:clpe}

We define a minimal language to express parallelism and interaction of LPEs;
the operators are taken from mCRL2~\cite{GrooteM2014} and similar-styled process algebras.
Let $\comms$ be the set of \emph{communication} expressions of the form $a_0 | \ldots | a_n \rightarrow c$ where $a_0, \ldots, a_n, c \in \events$ are action labels.

\begin{definition}
  The process algebra is defined as follows:
  \begin{equation*}
    S ::= \communication_{C}(S) ~\mid~ \allow_A(S) ~\mid~ \hide_H(S) ~\mid~ S \parallel S ~\mid~ P(\initval)
  \end{equation*}
  Here, $A \subseteq 2^{\events \rightarrow \naturalnumbers}$ is a non-empty finite set of finite multi-sets of action labels, $H \subseteq \events$ is a non-empty finite set of action labels and $C \subseteq \comms$ is a finite set of \emph{communications}.
  Finally, we have $P(\initval) \in \linearprocess$.
\end{definition}

The set $\simpleprocess$ contains all expressions of the process algebra.
Operator $\communication_{C}$ describes \emph{communication}, $\allow_A$ \emph{action allowing}, $\hide_H$ \emph{action hiding} and $\parallel$ \emph{parallel composition};
the elementary objects are the processes, defined as LPEs.

The operational semantics of expressions in $\simpleprocess$ are defined in Definition~\ref{def:operational_semantics}.
We first introduce three auxiliary functions on $\semactions$ that are used in the semantics.

\begin{definition}
  Given $\action \in \semactions$ we define $\actcomm{C}$, where $C \subseteq \comms$, as follows:
  \begin{align*}
    \actcomm{\emptyset}(\action) &= \action \\
    \actcomm{C}(\action) &= \actcomm{C \setminus C_1}(\actcomm{C_1}(\action))  \text{ for } C_1 \subset C \\
    \actcomm{\set{a_0 | \ldots | a_n \rightarrow c}}(\action) &=
      \begin{cases}
        \parbox{0.6\columnwidth}{$\multiset{c(\sem{d})} + \actcomm{\{a_0 | \ldots | a_n \rightarrow c\}}(\action - \multiset{a_0(\sem{d}), \ldots, a_n(\sem{d})})$} \\
        		    \quad\,\,\,\, \text{if } \multiset{a_0(\sem{d}), \ldots, a_n(\sem{d})} \subseteq \action \\
        \action \quad \text{otherwise}
      \end{cases}
  \end{align*}
\end{definition}

For $\actcomm{C}$ to be well-defined we require that the left-hand sides of the communications do not share labels.
Furthermore, the action label on the right-hand side must not occur in any \emph{other} left-hand side.
For example $\actcomm{\set{a|b\rightarrow c}}(a|d|b) = c|d$, but $\actcomm{\set{a|b\rightarrow c, a|d\rightarrow c}}(a|d|b)$ and $\actcomm{\set{a|b\rightarrow c,c\rightarrow d}}(a|d|b)$ are not allowed.

\begin{definition}
  Let $\omega \in \semactions$, $H \subseteq \events$ and $\omega \in \Omega$. We define $\acthide{H}(\action)$ as the multi-set $\omega'$ defined as:
  \begin{equation*}
  	\action'(a(\sem{d})) = 
  	\begin{cases}
  		0 &\text{if } a \in H \\
  		\action(a(\sem{d})) &\text{otherwise}
  	\end{cases}
  \end{equation*}
\end{definition}

Finally, given a multi-action $\multiact$ we define $\nodata{\multiact}$ to obtain the multi-set of action labels, \eg, $\nodata{a(3)|b(5)} = \multiset{a, b}$.
Formally, $\nodata{a(e)} = \multiset{a}$, $\nodata{\tau} = \multiset{}$ and $\nodata{\alpha|\beta} = \nodata{\alpha} + \nodata{\beta}$.
We define $\nodata{\action}$ for $\action \in \semactions$ in a similar way.
  
\begin{definition}\label{def:operational_semantics}
  The operational semantics of an expression $Q$ of $\simpleprocess$, denoted $\interpret{Q}$, are defined by the corresponding LTS $(\simpleprocess, Q, \semactions, \transitions)$ with its transition relation defined by the rules below and the transition relation given in Definition~\ref{def:lpesemantics} for each expression in $\linearprocess$.
  For any $\action,\action' \in \semactions$, expressions $P, P', Q, Q'$ of $\simpleprocess$ and sets $C \subseteq \comms$, $A \subseteq 2^{\events \rightarrow \naturalnumbers}$ and $H \subseteq \events$:
  
  \begin{center}
  \begin{tabular}{c c}
    $\inferrule*[Left=Com]{P \xrightarrow{\action} P'   }{\communication_{C}(P) \xrightarrow{\actcomm{C}(\action)} \communication_{C}(P')}$  &
    \quad\quad\quad $\inferrule*[Left=Allow,Right=$\nodata{\action} \in A$]{P \xrightarrow{\action} P'  }{\allow_A(P) \xrightarrow{\action} \allow_A(P')}$ \\[4ex]

    $\inferrule*[Left=Hide]{P \xrightarrow{\action} P' }{\hide_H(P) \xrightarrow{\acthide{H}(\action)} \hide_H(P')}$
    & $\inferrule*[Left=Par]{P \xrightarrow{\action} P' \quad Q \xrightarrow{\action'}  Q'}{P \parallel Q \xrightarrow{\action \,+\, \action'} P' \parallel Q'}$ \\[4ex]

    $\inferrule*[Left=ParR]{Q \xrightarrow{\action} Q'}{P \parallel Q \xrightarrow{\action} P \parallel Q'}$
    & $\inferrule*[Left=ParL]{P \xrightarrow{\action} P'}{P \parallel Q \xrightarrow{\action} P' \parallel Q}$  
  \end{tabular}
  \end{center}
  Note that for $\textsc{Allow}$ the condition $\nodata{\action} \in A$ must be satisfied in order for the rule to be applicable.
\end{definition}

\begin{example}
Consider the following LPE that models a drill component in which each $\acttoggle$ action leads to a $\actdrill$ action.
\begin{align*}
	\procdrill(t: \boolsort) = (\neg t) &\rightarrow \acttoggle \sequential \procdrill(\true) \\
	           +\quad(t) &\rightarrow \actdrill \sequential \procdrill(\false) 
\end{align*}
Suppose that we wish to study the interaction of LPEs $\procmachine$ of Example~\ref{example:delayedmachine} and $\procdrill$, assuming that their $\acttoggle$ actions must synchronise, resulting in a $\overline{\acttoggle}$ action.
Let $C =\{\acttoggle|\acttoggle \rightarrow \overline{\acttoggle}\}$ be the communication function that specifies this synchronisation, and
let $A = \{\multiset{\overline{\acttoggle}},\multiset{\actdrill},\multiset{\actcount}\}$ be the set of multi-action labels we allow.
The interaction between LPEs $\procmachine$ and $\procdrill$ can be specified by the expression
$\allow_A(\communication_C(\procmachine(0, \false) \parallel \procdrill(\false)))$
in the algebra.
An example derivation is depicted below:
{\begin{center}\footnotesize
$
\inferrule*[Left=Par]{\procmachine(0,\false) \transition{\multiset{\acttoggle}} \procmachine(3,\true) \qquad\qquad
                      \procdrill(\false) \transition{\multiset{\acttoggle}} \procdrill(\true)
                     }
                     {\inferrule*[Left=Com]{\procmachine(0, \false) \parallel \procdrill(\false) \transition{\multiset{\acttoggle:2}} \procmachine(3, \true) \parallel \procdrill(\true)}
                                           {\inferrule*[Left=Allow,Right=$\normalfont \nodata{\multiset{\overline{\acttoggle}}} \in A$]{\communication_C(\procmachine(0, \false) \parallel \procdrill(\false)) 
\transition{\multiset{\overline{\acttoggle}}} \communication_C(\procmachine(3, \true) \parallel \procdrill(\true))}
                                                                   {\allow_A(\communication_C(\procmachine(0, \false) \parallel \procdrill(\false)))
\transition{\multiset{\overline{\acttoggle}}} 
\allow_A( \communication_C(\procmachine(3, \true) \parallel \procdrill(\true)))
                                                                   }
                                           }
                     }
$
\end{center}
}
Note that we cannot derive a $\multiset{\acttoggle}$ transition for 
$\allow_A(\communication_C(\procmachine(0, \false) \parallel \procdrill(\false)))$, even though we can derive, \emph{e.g.},
$\procmachine(0, \false) \parallel \procdrill(\false) \transition{\multiset{\acttoggle}} \procmachine(3, \true) \parallel \procdrill(\false)$ by rule \textsc{ParL}.
The reason for this is that $\nodata{\multiset{\acttoggle}} \notin A$.
\end{example}

\section{The Decomposition Problem}\label{section:decomposition}

The state space of a monolithical LPE may grow quite large and generating that state space may either take too long or require too much memory.
We are therefore interested in decomposing an LPE into two or more LPEs, where the latter are referred to as \emph{components}, such that the state spaces of the resulting components are smaller than that of the original state space. 
Such a decomposition is considered \emph{valid} iff the original state space is strongly bisimilar to the state space of these components when combined under a suitable context (\emph{i.e.}, an expression with a `hole') which formalises how to combine the components.
We formalise this problem as follows.

\begin{definition}\label{def:valid_decomposition}
  Let $P(\vec{d} : \vec{D}) = \phi$ be an LPE and $\vec{\initval} : \vec{D}$ a closed expression.
  The LPEs $P_0(\subvector{\vec{d}}{I_0} : \subvector{\vec{D}}{I_0}) = \phi_0$ to $P_n(\subvector{\vec{d}}{I_n} : \subvector{\vec{D}}{I_n}) = \phi_n$, for indices $I_0, \ldots, I_n \subseteq \naturalnumbers$, are a valid \emph{decomposition} of $P$ and $\vec{\initval}$ iff there is a context $\mathsf{C}$ such that:  
  \begin{equation*}
    \interpret{P(\vec{\initval})} \bisim \interpret{\mathsf{C}[P_0(\subvector{\vec{\initval}}{I_0}) \parallel \ldots \parallel P_n(\subvector{\vec{\initval}}{I_n})]}
  \end{equation*}
  Where $\mathsf{C}[P_0(\subvector{\vec{\initval}}{I_0}) \parallel \ldots \parallel P_n(\subvector{\vec{\initval}}{I_n})]$ is an expression in $\simpleprocess$.
  We refer to the expression $\mathsf{C}[P_0(\subvector{\vec{\initval}}{I_0}) \parallel \ldots \parallel P_n(\subvector{\vec{\initval}}{I_n})]$ as the \emph{composition}.
\end{definition}

In the next sections, we will show that a suitable context $\mathsf{C}$ can be constructed using the operators from $\simpleprocess$, and
we define a decomposition technique that results in exactly two components (a \emph{cleave}). 
The technique can, in principle, be applied recursively to the smaller components.
The primary benefit of a valid decomposition is that a state space that is equivalent to the original state space can be obtained as follows.
First, the state space of each component is derived separately.
The composition can then be derived from the component state spaces, exploiting the rules of the operational semantics.
The component state spaces can be minimised modulo bisimilarity, which is a congruence with respect to the operators of $\simpleprocess$ before deriving the results of the composition expression.
The composition resulting from these minimised components can be considerably smaller than the original state space, because also the original state space can often be reduced considerably modulo strong bisimilarity after generation.
This process is referred to as \emph{compositional minimisation}.

\section{A Solution to the Decomposition Problem}\label{section:solution}

A basic observation that we exploit in our solution to the decomposition problem is that when hiding label $c$ in a multi-action $\multiact | c$, we are left with multi-action $\multiact$, provided that $c$ does not occur in $\multiact$.
When the multi-action $\multiact$ is an event that is possible in a monolithic LPE and the label $c$ is the result of a communication of two components, we can effectively exchange information between multiple components, without this information becoming visible externally.
The example below illustrates the idea using a naive but valid solution to the decomposition technique on the LPE of Example~\ref{example:delayedmachine}.

\begin{example}\label{example:naivecleave}
	Reconsider the LPE \emph{$\procmachine$} we defined earlier, and consider the two components depicted below.
	\begin{align*}
		\procmachine_V(n : \natsort)
		    & = \sum_{s : \boolsort} (n > 0) \rightarrow  \actcount | \actsync^0_V(n, s) \sequential \procmachine_V(n - 1) \\
		    & + \sum_{s : \boolsort} (n \approx 0) \rightarrow \actsync^1_V(n, s) \sequential \procmachine_V(\textsf{if}(\neg s, 3, 1)) \\
 		\procmachine_W(s : \boolsort)
		    & = \sum_{n : \natsort} (n > 0) \rightarrow \actsync^0_W(n, s) \sequential \procmachine_W(s) \\
		    & + \sum_{n : \natsort} (n \approx 0) \rightarrow \acttoggle | \actsync^1_W(n, s) \sequential \procmachine_W(\neg s)
	\end{align*}
  Each component describes part of the behaviour and knows the value of parameter $n$ or $s$, but not the other.
  To cater for this, it is `over-approximated' by a sum variable.
  The state space of $\procmachine_V(0)$ is shown below.
  The synchronisation actions $\actsync$ expose the non-deterministically chosen values of the unknown parameters.
  
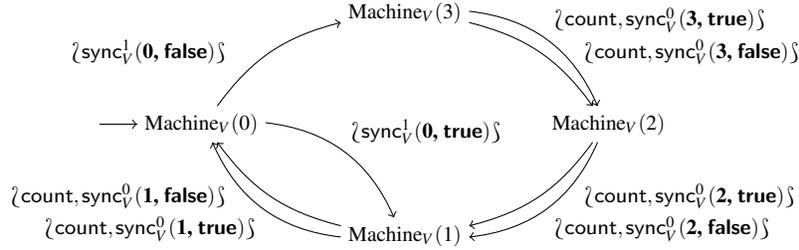
\begin{center}
 	\begin{tikzpicture}[->]
 		\scriptsize
	  \node (s0) {$\procmachine_V(0)$};
	  \node (s0init) [left= 0.5cm of s0] {};
	  \node (s1) [above right= of s0] {$\procmachine_V(3)$};
	  \node (s2) [below right= of s1] {$\procmachine_V(2)$};
	  \node (s3) [below right= of s0] {$\procmachine_V(1)$};
 	
 		\draw (s0init) edge (s0);
	  \draw (s0) edge[bend left] node[above right] {$\multiset{\actsync^1_V(\textbf{0, \true})}$} (s3);
	  \draw (s0) edge[bend left=20] node[left=.5cm] {$\multiset{\actsync^1_V(\textbf{0, \false})}$} (s1);
	  
	  \draw (s1) edge[bend left] node[above right] {$\multiset{\actcount,\actsync^0_V(\textbf{3, \true})}$} (s2);	  
	  \draw (s1) edge[bend left=20] node[right=0.5cm] {$\multiset{\actcount,\actsync^0_V(\textbf{3, \false})}$} (s2);
	  
 	  \draw (s2) edge[bend left] node[below right] {$\multiset{\actcount,\actsync^0_V(\textbf{2, \false})}$} (s3);
	  \draw (s2) edge[bend left=20] node[right=.5cm] {$\multiset{\actcount,\actsync^0_V(\textbf{2, \true})}$} (s3);
	  
 	  \draw (s3) edge[bend left] node[below left] {$\multiset{\actcount,\actsync^0_V(\textbf{1, \true})}$} (s0);
 	  \draw (s3) edge[bend left=20] node[left=.5cm] {$\multiset{\actcount,\actsync^0_V(\textbf{1, \false})}$} (s0);
 	\end{tikzpicture}
  \end{center}  
  Enforcing synchronisation of the $\actsync$ actions, the context $\mathsf{C}$ can be chosen as follows to achieve a valid decomposition:
  \begin{align*}
    & \allow_{\{\multiset{\acttoggle}, \multiset{\actcount}\}}( \hide_{\{\actsync^0, \actsync^1\}}( \communication_{\{\actsync^0_V|\actsync^0_W \rightarrow \actsync^0, \actsync^1_V|\actsync^1_W \rightarrow \actsync^1\}}(\procmachine_V(0) \parallel \procmachine_W(\false))))
  \end{align*}
\end{example}
  
Unfortunately the state space of $\procmachine_W(\false)$ in the above example is infinitely branching and it has no finite state space that is strongly bisimilar to it, rendering the decomposition useless in practice.
We will subsequently develop a more robust solution.
\subsection{Separation Tuples}

To obtain a useful decomposition it can be beneficial to reduce the number of parameters that occur in the synchronisation actions, because these then become a visible part of the transitions in the state spaces of the individual components.
In the worst case, as illustrated by LPE $\procmachine_W$ of Example~\ref{example:naivecleave}, synchronisation actions lead to a component having an infinite state space despite the fact that the state space of the original LPE is finite.

One observation we exploit is that in some cases we can actually remove the synchronisation for summands completely.
For instance, in the first summand of $\procmachine$ in Example~\ref{example:delayedmachine}, the value of parameter $s$ remains unchanged and the condition is only an expression containing parameter $n$.
We refer to summands with such a property as \emph{independent} summands.
When defining the context $\mathsf{C}$, we can allow a component to execute multi-actions of its independent summands without enforcing a synchronisation with the other component.
This allows, for instance,  component $\procmachine_V$ to independently execute (multi-)action $\actcount$ without synchronising the values of $s$ and $n$ with $\procmachine_W$.

A second observation that we exploit is that if there are independent summands, then not every summand needs to be present in both components.
However, we must ensure that each summand of the monolithic LPE is covered by at least one of the two components that we extract from the LPE.
The summands that we extract for a given component are identified by a set of indices $\Isub$ of the summands of the monolithic LPE.
Of these, we furthermore can identify summands that are dependent and summands that are independent.
The indices for the latter are collected in the set $\Iind$.

A third observation that can be utilised is that for the dependent summands, there is some degree of flexibility for deciding which part of the summand of the monolithic LPE will be contributed by which component.
More specifically, by carefully distributing the enabling condition $c$ and action expression $\multiact$ of a summand of the monolithic LPE over the two components, the amount of information (\emph{i.e.}, information about `missing' parameters, given by a synchronisation expression $h$) that needs to be exchanged between these two components when they execute their respective summands, can be minimised.

Note that the way we distribute the list of process parameters of the monolithic LPE over the two components may affect which summands can be considered independent.
For instance, had we decided to assign the (multi-)action $\actcount$ to $\procmachine_W$ and $\acttoggle$ to $\procmachine_V$, we would not be able to declare $\actcount$'s summand independent.
Consequently, the set of process parameter indices $U$, assigned to a component, and the set $K$ are mutually dependent.
To capture this relation, we introduce the concept of a \emph{separation tuple}.
The concept of a separation tuple, a 6-tuple which we introduce below, formalises the required relation between $\Iind$, $\Isub$ and $U$, and the conditions $c$, and action $\multiact$ and synchronisation expressions $h$ of a component.
To define the expressions we use indexed sets where the index of each element, indicated by a subscript, determines the index of the summand to which the expression belongs.

\begin{definition}\label{def:derived}
  Let $P(\vec{d} : \vec{D}) = \bigplus_{i \in I} \sum_{e_i : E_i} c_i \rightarrow \multiact_i \sequential P(\vec{g}_i)$ be an LPE.
  A \emph{separation tuple} for $P$ is a 6-tuple $(U, \Iind, \Isub, c^U, \multiact^U, h^U)$ where $U \subseteq \naturalnumbers$ is a set of parameter indices, $\Iind \subseteq \Isub \subseteq I$ are two sets of summand indices, and
  $c^U, \multiact^U$ and $h^U$ are indexed sets of condition, action and synchronisation expressions respectively.
We require that for all $i \in (\Isub \setminus \Iind)$ it holds that $\freevars(c^U_i) \cup \freevars(\multiact^U_i) \cup \freevars(\vec{h}^U_i) \subseteq \flatten{\vec{d}} \cup \{e_i\}$,
  and for all $i \in K$ it holds that $\freevars(c_i) \cup \freevars(\multiact_i) \cup \freevars(\subvector{\vec{g_i}}{U}) \subseteq \flatten{\subvector{\vec{d}}{U}} \cup \{e_i\}$.
   
  A separation tuple induces an LPE, where $U^c = \naturalnumbers \setminus U$, as follows:
  \begin{align*}
    P_{U}(\subvector{\vec{d}}{U} : \subvector{\vec{D}}{U}) = &\bigplus_{i \in (\Isub \setminus \Iind)}~ \sum_{e_i : E_i, \subvector{\vec{d}}{U^c} : \subvector{\vec{D}}{U^c}} 
   \hspace{-1.5em}c^U_i \rightarrow \multiact^U_i|\textsf{sync}^i_U(\vec{h}^U_i) \sequential P_U(\subvector{\vec{g_i}}{U}) \\
  + &~~ \bigplus_{i \in K}~ \sum_{e_i : E_i} c_i \rightarrow \multiact_i | \textsf{tag} \sequential P_U(\subvector{\vec{g_i}}{U})
  \end{align*}
  We assume that action label $\actsync^i_U$, for any $i \in I$, and label $\acttag$ does not occur in $\multiact_j$, for any $j \in I$, to ensure that these action labels are fresh.
\end{definition}

Observe that for independent summands the action label is extended with a $\acttag$ action in Definition~\ref{def:derived}.
This label is needed to properly deal with overlapping multi-actions, as we illustrate below in Example~\ref{ex:roletag}.
\begin{example}\label{ex:roletag}
Consider the following LPE.
\begin{align*}
	\text{P}(x : \boolsort, y : \boolsort) &= x  \rightarrow a \sequential \text{P}(\false, y) \\
																	&+ y \rightarrow b \sequential \text{P}(x, \false) \\
																	&+ (x \land \neg y) \rightarrow a|b \sequential \text{P}(\false, \false)
\end{align*}
Suppose we decompose LPE $P$ using the separation tuple $(V, \{0\}, \{0,2\}, \{x_2\}, \{a_2\}, \{\vector{}_2\})$ and the tuple $(W, \{1\}, \{1,2\}, \{(\neg y)_2\}, \{b_2\}, \{\vector{}_2\})$, where $V = \{0\}$ and $W = \{1\}$.
Now assume that we had omitted the $\acttag$ action in Definition~\ref{def:derived}, in which case 
these separation tuples would induce the following LPEs:
\begin{align*}
	P_V(x : \boolsort) &= x \rightarrow a \sequential P_V(\false) \\
																	  &+ x \rightarrow a|\actsync_V^2 \sequential P_V(\false) \\
 	P_W(y : \boolsort) &= y \rightarrow b \sequential P_W(\false) \\
  &+ (\neg y) \rightarrow b|\actsync_W^2 \sequential P_W(\false)
\end{align*}
Now, observe both $P_V(\true) \transition{\multiset{a}} P_V(\false)$ and $P_W(\true) \transition{\multiset{b}} P_W(\false)$ are transitions for these components.
This also means that due to (among others) rule $\textsc{Par}$, $P_V(\true) \parallel P_W(\true)$ can perform action $\multiset{a,b}$.
Note that process $P(\true, \true) $ does \emph{not} have an outgoing transition labelled with $\multiset{a,b}$, but (the reachable) process $P(\true,\false)$ \emph{does} have an outgoing $\multiset{a,b}$ transition.
There is, however, no composition expression that prevents $\multiset{a,b}$ in $P_V(\true) \parallel P_W(\true)$ and allows $\multiset{a,b}$ in $P_V(\true) \parallel P_W(\false)$.
The $\acttag$ label provides the tools for making this distinction.
\end{example}

The components, induced by two separation tuples, can be (re)combined in a context that enforces synchronisation of the $\actsync$ events and which hides their communication trace. This ensures that all actions left can be traced back to the monolithic LPE from which the components are derived.
Under specific conditions, this is achieved by the following context.
\begin{definition}\label{def:refinedcleavecomposition}
  Let $P(\vec{d} : \vec{D}) = \bigplus_{i \in I} \sum_{e_i : E_i} c_i \rightarrow \multiact_i \sequential P(\vec{g}_i)$ be an LPE and $(V, \Iind^V, \Isub^V, c^V, \multiact^V, h^V)$ and $(W, \Iind^W, \Isub^W, c^W, \multiact^W, h^W)$ be separation tuples for $P$.
  Let $P_V(\subvector{\vec{d}}{V} : \subvector{\vec{D}}{V}) = \phi_V$ and $P_W(\subvector{\vec{d}}{W} : \subvector{\vec{D}}{W}) = \phi_W$ be the induced LPEs according to Definition~\ref{def:derived}.
  Let $\initval : \vec{D}$ be a closed expression.
  Then the composition expression is defined as:
  \begin{align*}
    \hide_{\{\acttag\}} (
    	\allow_{\{\nodata{\multiact_i} \mid i \in I\} \cup \{\nodata{\multiact_i|\acttag} \mid i \in (\Iind^V \cup \Iind^W)\}}( 
        \hide_{\{\actsync^i \mid i \in I\}} (\communication_{\{ \actsync^i_V| \actsync^i_W \rightarrow\,\actsync^i \mid i \in I\}} (P_V(\subvector{\vec{\initval}}{V}) \parallel P_W(\subvector{\vec{\initval}}{W})))
    	  )
      )
  \end{align*}
\end{definition}

Before we proceed to identify the conditions under which two separation tuples induce a valid decomposition using the above context,
we revisit Example~\ref{example:delayedmachine} to illustrate the concepts introduced so far.

\begin{example}\label{example:refinedcleave}
	Reconsider the LPE presented in Example~\ref{example:delayedmachine} with $V = \{0\}$ and $W = \{1\}$.
	The separation tuple $(V, \{0\}, \{0,1\}, \{(n \approx 0)_1\}, \{\tau_1\}, \{\vector{s}_1\})$ 
  and $(W, \emptyset, \{1\}, \{\true_1\}, \{\acttoggle_1\},$ $\{\vector{s}_1\})$ for $\procmachine$ induce component $\procmachine_V$ and $\procmachine_W$ respectively.
  \begin{align*}
    \procmachine_V(n : \natsort) &= \quad\,\,\,\,\,\, (n > 0) \rightarrow \actcount|\acttag \sequential \procmachine_V(n - 1) \\
      &+ \sum_{s : \boolsort} (n \approx 0) \rightarrow \tau | \actsync^1_V(s) \sequential \procmachine_V(\textsf{if}(\neg s, 3, 1)) \\
    \procmachine_W(s : \boolsort) &= \true \rightarrow \acttoggle | \actsync^1_W(s) \sequential \procmachine_W(\neg s)
  \end{align*}
  Note that we omitted the $\sum$-operator in the first summand of $\procmachine_V$ since sum variable $s$ does not occur as a free variable in the expressions; for similar reasons, the summand of $\procmachine_W$ is omitted.
  The state spaces of components $\procmachine_V(0)$ and $\procmachine_W(\false)$ are shown in Figure~\ref{figure:decomposition}.
  We obtain the following composition according to Definition~\ref{def:refinedcleavecomposition}:
  \begin{align*}
  	\hide_{\{\acttag\}}(&\allow_{\{\multiset{\acttoggle},\multiset{\actcount},\multiset{\actcount,\acttag}\}} (\hide_{\{\actsync^0,\actsync^1\}} (\\
  	&\communication_{
  		\{\actsync^0_V|\actsync^0_W \rightarrow \actsync^0, \actsync^1_V|\actsync^1_W \rightarrow \actsync^1\}
  		} (\procmachine_V(0) \parallel \procmachine_W(\false)))))
  \end{align*}
  The state space of this expression is strongly bisimilar to the state space of $\procmachine(0, \false)$ shown in Figure~\ref{figure:delayedmachine}.
  Note that the state space of $\procmachine_V(0)$ has four states and transitions, and the state space of $\procmachine_W(\false)$ has two states and transitions, which are both smaller than the original state space.
  Their composition has the same size as the original state space and no further minimisation can be achieved (note that the state space of Figure~\ref{figure:delayedmachine} is already minimal).
\end{example}

\subsection{Cleave Correctness Criteria}

It may be clear that not every decomposition which satisfies Definition~\ref{def:refinedcleavecomposition} yields a \emph{valid} decomposition (in the sense of Definition~\ref{def:valid_decomposition}).
For example, replacing the condition expression $\true$ in Example~\ref{example:refinedcleave} of the summand in $P_W$ by $\false$ would not result in a valid decomposition.
Our aim in this section is to present the necessary and sufficient conditions to establish that the state space of the monolithic LPE is bisimilar to the state space of the composition expression resulting from Definition~\ref{def:refinedcleavecomposition}.

Consider a decomposition of an LPE $P$ according to Definition~\ref{def:refinedcleavecomposition}, induced by separation tuples $(V, \Iind^V, \Isub^V, c^V, \multiact^V, h^V)$ and $(W, \Iind^W, \Isub^W, c^W, \multiact^W, h^W)$.
We abbreviate the composition expression of Definition~\ref{def:refinedcleavecomposition} by $\mathsf{C}[P_V(\subvector{\vec{d}}{V}) || P_W(\subvector{\vec{d}}{W})]$.
Recall that components $P_V$ and $P_W$ yield a valid decomposition of $P$ if there is a bisimulation relation between $P(\vec{d})$ and $\mathsf{C}[P_V(\subvector{\vec{d}}{V}) || P_W(\subvector{\vec{d}}{W})]$.
Bisimilarity requires that states that are related can mimic each other's steps.
Since three LPEs are involved (the LPE $P$ and the two interacting components, induced by the separation tuples), we must consider situations that can emerge from any of these three LPEs executing a (multi-)action.

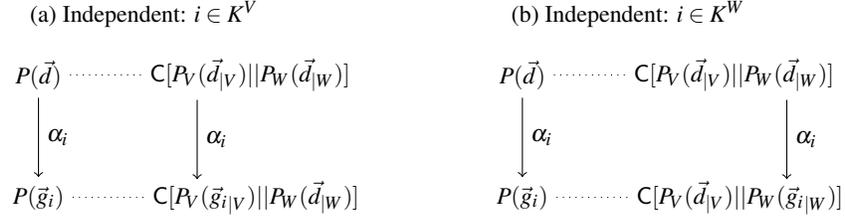
\begin{figure}[h]
\footnotesize\centering
\begin{tikzpicture}
\node[draw=none] (P1) {$P(\vec{d})$};
\node[draw=none, right=27pt of P1] (C1) {$\mathsf{C}[P_V(\subvector{\vec{d}}{V}) || P_W(\subvector{\vec{d}}{W})]$};
\node[draw=none, below=30pt of P1] (P1') {$P(\vec{g}_i)$};
\node[draw=none, right=27pt of P1'] (C1') {$\mathsf{C}[P_V(\subvector{\vec{g_i}}{V}) || P_W(\subvector{\vec{d}}{W})]$};
\coordinate (MiddleP1C1) at ($(P1)!0.5!(C1)$);
\node[draw=none, above=15pt of MiddleP1C1] (case1) {(a) Independent: $i \in K^V$};


\node[draw=none, right=50pt of C1] (P3) {$P(\vec{d})$};
\node[draw=none, right=27pt of P3] (C3) {$\mathsf{C}[P_V(\subvector{\vec{d}}{V}) || P_W(\subvector{\vec{d}}{W})]$};
\node[draw=none, below=30pt of P3] (P3') {$P(\vec{g}_i)$};
\node[draw=none, right=27pt of P3'] (C3') {$\mathsf{C}[P_V(\subvector{\vec{d}}{V}) || P_W(\subvector{\vec{g_i}}{W})]$};
\coordinate (MiddleP3C3) at ($(P3)!0.5!(C3)$);
\node[draw=none, above=15pt of MiddleP3C3] (case3) {(b) Independent: $i \in K^W$};

\path[->]
  (P1) edge node[right] {$\alpha_i$} (P1')
  (P3) edge node[right] {$\alpha_i$} (P3')
;
\draw
  ([xshift=-20pt]C1.south) edge[->] node[right] {$\alpha_i$} ([xshift=-20pt,yshift=-30pt]C1.south)
  ([xshift=20pt]C3.south) edge[->] node[right] {$\alpha_i$} ([xshift=20pt,yshift=-30pt]C3.south)
;
\path[dotted]
  (P1) edge (C1)
  (P1') edge (C1')
  (P3) edge (C3)
  (P3') edge (C3')
;

\end{tikzpicture}

\caption{Two of the possible situations that must be considered when showing the validity of the decomposition of Definition~\ref{def:refinedcleavecomposition}: the execution of independent summands depicted in situations (a) and~(b).}
\label{fig:possibilities}
\end{figure}

Two of the three relevant scenarios that must be considered are depicted in Figure~\ref{fig:possibilities}. 
Note that in all relevant scenarios, the initiative of the transition may be with either $P(\vec{d})$, or with the composition $\mathsf{C}[P_V(\subvector{\vec{d}}{V}) || P_W(\subvector{\vec{d}}{W})]$.

Suppose that the monolithic LPE $P$ can take a step due to some summand $i \in I$, for which also $i \in \Iind^V$.
In that case---case (a) in Figure~\ref{fig:possibilities}---Definition~\ref{def:refinedcleavecomposition} guarantees that the free variables of their condition, action and update expressions are taken from $\subvector{\vec{d}}{V}$; (multi-)action $\alpha_i$ matches (multi-)action $\alpha_i|\acttag$ after hiding $\acttag$.
However, this is not sufficient to guarantee full independence of both components: what may happen is that the execution of a summand that is assumed to be independent still modifies the value of a process parameter of the other component, violating the idea of independence, and resulting in a target state in the composition that cannot be related to the target state of the monolithic LPE.
In order to guarantee true independence, we must require that the $W$-projection on the update expression $\vec{g}_i$ of $P$ does not modify the corresponding parameters.
Case~(b) in Figure~\ref{fig:possibilities} is dual.
Formally, we require (\REQ{2}): for all $r \in \Iind^V$ we have $\subvector{\vec{g_r}}{W} = \subvector{\vec{d}}{W}$ and for all $r \in \Iind^W$ we demand $\subvector{\vec{g_r}}{V} = \subvector{\vec{d}}{V}$.
Note that in case $\Iind^V$ and $\Iind^W$ overlap, condition (\REQ{2}) guarantees that the involved summands only induce self-loops.
Finally, observe that (\REQ{2}) is also a sufficient condition for the monolithic LPE $P$ to match a (multi-)action $\alpha_i$ due to $P_V$ or $P_W$ in either of these two cases.

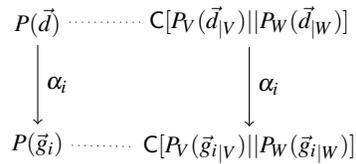
\begin{figure}[h]
\footnotesize\centering
\begin{tikzpicture}
\node[draw=none] (P2) {$P(\vec{d})$};
\node[draw=none, right=27pt of P2] (C2) {$\mathsf{C}[P_V(\subvector{\vec{d}}{V}) || P_W(\subvector{\vec{d}}{W})]$};
\node[draw=none, below=30pt of P2] (P2') {$P(\vec{g}_i)$};
\node[draw=none, below=30pt of C2] (C2') {$\mathsf{C}[P_V(\subvector{\vec{g_i}}{V}) || P_W(\subvector{\vec{g_i}}{W})]$};

\path[->]
  (P2) edge node[right] {$\alpha_i$} (P2')
;
\draw
  (C2) edge[->] node[right] {$\alpha_i$} (C2')
;
\path[dotted]
  (P2) edge (C2)
  (P2') edge (C2')
;

\end{tikzpicture}

\caption{The third possible situation that must be considered when showing the validity of the decomposition of Definition~\ref{def:refinedcleavecomposition}: the synchronous execution of summands.}
\label{fig:possibilities2}
\end{figure}

The more complex scenario that must be considered is when $P_V$ and $P_W$ (must) synchronise to mimic the behaviour of $P$; see Figure~\ref{fig:possibilities2}.
Suppose again that the monolithic LPE $P$ can execute an $\alpha_i$ action due to summand $i \in I$, but in this case, neither $i \in \Iind^V$, nor $i \in \Iind^W$.
First, observe that the only option to match the behaviour of this summand is if a component covers at least all summands not already covered by the other component.
We must therefore require at least the following (\REQ{1}): $\Isub^V = I \setminus \Iind^{W}$ and $\Isub^W = I \setminus \Iind^{V}$.



Second, observe that the enabledness of summand $i$ in $P$ depends on the enabling condition $c_i$.
Consequently, if $c_i$ holds true, then the $i$-indexed conditions $c_i^V$ and $c_i^W$ must also hold true. 
Moreover, since we are dealing with dependent summands, the multi-action expression $\multiact^V_i | \multiact^W_i$ must reduce to $\multiact_i$ under these conditions.
Also the additional synchronisation vectors $\vec{h}^V$ and $\vec{h}^W$ must agree, for otherwise the $\actsync$ actions of both components cannot participate in the synchronisation.
Note that we do not need to explicitly require relating the update expressions of $P$ and the components $P_V$ and $P_W$ resulting from the execution of their $i$-indexed summands, since this property is already guaranteed by construction; see Definition~\ref{def:derived}.
We collectively refer to the above requirements by condition (\REQ{3}).

\emph{Vice versa}, whenever both components can simultaneously execute their $i$-indexed summand, we must ensure that also the monolithic LPE $P$ can execute its $i$-indexed summand.
Condition (\REQ{4}) ensures that this requirement is met.  
Note that $P_V$ and $P_W$ only synchronise on summands with equal indices due to the synchronisation on $\actsync$ actions that is enforced.
A technical complication in formalising requirement (\REQ{4}), however, is that the sum variables of the individual components carry the same name in all three LPEs.
In particular, from the fact that both individual components can successfully synchronise, we cannot deduce a unique value assigned to these homonymous sum-variables.
We must therefore also ensure that the update expressions of the components, resulting from the executing the $r$-indexed summands, indeed is the same as could have resulted from executing the $r$-indexed summand in $P$.
Contrary to requirement (\REQ{3}), this property is not guaranteed by the construction of Definition~\ref{def:derived}, so there is a need to explicitly require it to hold.

A pair of separation tuples of $P$ satisfying the above requirements is called a \emph{cleave} of $P$.
Below, we formalise this notion, together with the requirements we informally introduced above.

\begin{definition}\label{def:requirements}
  Let $P(\vec{d} : \vec{D}) = \bigplus_{i \in I} \sum_{e_i : E_i} c_i \rightarrow \multiact_i \sequential P(\vec{g}_i)$ be an LPE and $(V, \Iind^V, \Isub^V, c^V, \multiact^V, h^V)$ and $(W, \Iind^W, \Isub^W, c^W, \multiact^W, h^W)$ be separation tuples for $P$ as defined in Definition~\ref{def:derived}.
	The two separation tuples are a \emph{cleave} of $P$ iff the following requirements hold.
	
	\begin{enumerate}
		\item[\REQ{1}.] $\Isub^V = I \setminus \Iind^W$ and $\Isub^W = I \setminus \Iind^V$.
		
		\item[\REQ{2}.] For all $r \in \Iind^V$, $\subvector{\vec{g_r}}{W} = \subvector{\vec{d}}{W}$, and for all $r \in \Iind^W$, $\subvector{\vec{g_r}}{V} = \subvector{\vec{d}}{V}$.
		
		\item[\REQ{3}.] For all $r \in (\Isub^V \cap \Isub^W)$ and substitutions $\subst$ satisfying $\interpret{\subst(c_r)}$, also:
		  \begin{itemize}
		    \item $\interpret{\subst(c^V_r )}$ and $\interpret{\subst(c^W_r)}$, and
		
		    \item $\interpret{\subst(\vec{h}^V_r)} = \interpret{\subst(\vec{h}^W_r))}$, and
		
		    \item $\interpret{\subst(\multiact^V_r | \multiact^W_r)} = \interpret{\subst(\multiact_r)}$.
		  \end{itemize}
		  
		\item[\REQ{4}.] For all $r \in (\Isub^V \cap \Isub^W)$ and substitutions $\subst$ and $\subst'$ satisfying 
		  $\interpret{\subst(c^V_r)}$ and $\interpret{\sigma'(c^W_r)}$ and $\interpret{\subst(\vec{h}^V_r)} = \interpret{\sigma'(\vec{h}^W_r)}$, there is a substitution $\rho$ such that $\interpret{\rho(\subvector{\vec{d}}{V})} = \interpret{\sigma(\subvector{\vec{d}}{V})}$ and $\interpret{\rho(\subvector{\vec{d}}{W})} = \interpret{\sigma'(\subvector{\vec{d}}{W})}$ and:
      
      \begin{itemize}
        \item $\interpret{\rho(c_r)}$, and
	
        \item $\interpret{\subst(\multiact^V_r)|\subst'(\multiact^W_r)} = \interpret{\rho(\multiact_r)}$, and
	
        \item $\interpret{\subst(\subvector{\vec{g_r}}{V})} = \interpret{\rho(\subvector{\vec{g_r}}{V})}$, and
        
        \item $\interpret{\subst'(\subvector{\vec{g_r}}{W})} = \interpret{\rho(\subvector{\vec{g_r}}{W})}$.
      \end{itemize}
	\end{enumerate}
\end{definition}

\begin{example}\label{example:machinecleave}
We argue that the separation tuples inducing the decomposition obtained in Example~\ref{example:refinedcleave} are a cleave indeed.
First of all, the requirements $\REQ{1}$ and $\REQ{2}$ can be checked quite easily.
The requirements $\REQ{3}$ and $\REQ{4}$ both have to be checked for the summand with index one.
Consider the requirement $\REQ{3}$ with a substitution $\sigma$ assigning any value to $n$ (and any value to other variables due to totality) such that $\interpret{\sigma(n \approx 0)}$ holds.
It follows directly that both$\interpret{\sigma(n \approx 0)}$ and $\interpret{\sigma(\true)}$ hold.
Furthermore, $\interpret{\subst(\vector{s})} = \interpret{\subst(\vector{s})}$ and $\interpret{\subst(\tau | \acttoggle)} = \interpret{\subst(\acttoggle)}$ by definition.
For the requirement  $\REQ{4}$ consider any two substitutions $\sigma$ and $\sigma'$ such that both $\interpret{\sigma(n \approx 0)}$ and  $\interpret{\sigma'(\true)}$ hold and $\interpret{\sigma(\vector{s})} = \interpret{\sigma'(\vector{s})}$.
For substitution $\rho$ we can choose $n$ to be zero and $s$ to be equal to $\interpret{\sigma(\vector{s})}$.
The most interesting observation is that then indeed $\interpret{\sigma(\textsf{if}(\neg s, 3, 1))} = \interpret{\rho(\textsf{if}(\neg s, 3, 1))}$ and that $\interpret{\sigma'(\neg s)} = \interpret{\rho(\neg s)}$.
The other conditions are also satisfied and thus this is a cleave.
We can also observe that leaving out the synchronisation of $s$ does not yield a cleave since there is no substitution $\rho$ meeting the conditions in COM when substitutions $\sigma$ and $\sigma'$ disagree on the value of $s$.
\end{example}

Informally, we have already argued that the decomposition  yields a state space that is bisimilar to the original monolithic LPE.

We finish this section with a formal claim stating that a cleave induces a valid decomposition of a monolithic LPE.
The complete proof can be found in the technical report~\cite{Laveaux20}.

\begin{restatable}{theorem}{cleavecorrectness}\label{theorem:refinedcleave}
  Let $P(\vec{d} : \vec{D}) = \bigplus_{i \in I} \sum_{e_i : E_i} c_i \rightarrow \multiact_i \sequential P(\vec{g}_i)$ be an LPE and let
  $(V, \Iind^V, \Isub^V, c^V, \multiact^V, h^V)$ and $(W, \Iind^W, $ $\Isub^W, c^W, \multiact^W, h^W)$ be a cleave as defined in Definition~\ref{def:requirements}.
	For a closed expression $\vec{\initval} : \vec{D}$ the interpretation of the composition expression defined in Definition~\ref{def:refinedcleavecomposition} is strongly bisimilar to $\interpret{P(\vec{\initval})}$ and as such a valid decomposition according to Definition~\ref{def:valid_decomposition}.
\end{restatable}

\section{State Invariants}\label{section:state_invariant}

The separation tuples inducing the decomposition obtained in Example~\ref{example:refinedcleave} are indeed a cleave as shown in Example~\ref{example:machinecleave}, but this is by no means the only cleave for $\procmachine$.
For instance, also the decomposition we obtained in Example~\ref{example:naivecleave} can be achieved by means of a cleave.
The infinite branching of $\procmachine_W(\false)$ in that example is, however, problematic for the purpose of compositional minimisation.
While in this case, as shown by Example~\ref{example:refinedcleave}, we could avoid the infinite branching of $\procmachine_W(\false)$ by reducing the amount of synchronisation, this might not always be possible.

Another way to restrict the behaviour of the components is to strengthen the condition expressions of each summand, thus limiting the number of outgoing transitions.
We show that so-called \emph{state invariants}~\cite{BezemG94} can be used for this purpose.
These state invariants are typically formulated by the user based on the understanding of the model behaviour.

\begin{definition}
	Given an LPE $P(d : D) = \bigplus_{i \in I} \sum_{e_i : E_i} c_i \rightarrow \multiact_i \sequential P(g_i)$.
	A boolean expression $\psi$ such that $\freevars(\psi) \subseteq \{d\}$ is called a \emph{state invariant} iff the following holds:
	for all $i \in I$ and closed expressions $\initval : D$ and $l : E_i$ such that $\interpret{[d \gets \initval, e_i \gets l](c_i \land \psi)}$ holds then $\interpret{[d \gets \initval, e_i \gets l](g_i)](\psi)}$ holds as well.
\end{definition}

The essential property of a state invariant is that whenever it holds for the initial state it is guaranteed to hold for all reachable states in the state space.
This follows relatively straightforward from its definition.
Next, we define a \emph{restricted} LPE where (some of) the condition expressions are strengthened with a boolean expression.

\begin{definition}\label{def:restricted_lpe}
	Given an LPE $P(d : D) = \bigplus_{i \in I} \sum_{e_i : E_i} c_i \rightarrow \multiact_i \sequential P(g_i)$, a boolean expression $\psi$ such that $\freevars(\psi) \subseteq \{d\}$ and a set of indices $J \subseteq I$.
	We define the restricted LPE, denoted by $P^{\psi,J}$, as follows:  
  \begin{align*}
    P^{\psi,J}(d : D) = &\bigplus_{i \in J} \sum_{e_i : E_i} c_i \land \psi \rightarrow \multiact_i \sequential P^{\psi,J}(g_i) \\
    &+ \bigplus_{i \in (I \setminus J)} \sum_{e_i : E_i} c_i \rightarrow \multiact_i \sequential P^{\psi,J}(g_i)
  \end{align*}
\end{definition}

Note that if the boolean expression $\psi$ in Definition~\ref{def:restricted_lpe} is a state invariant for the given LPE then for all closed expressions $\vec{\initval} : \vec{D}$ such that $\interpret{[\vec{d} \gets \vec{\initval}](\psi)}$ holds, it holds that $\interpret{P(\vec{\initval})} \bisim \interpret{P^{\psi,J}(\vec{\initval})}$, for any $J \subseteq I$.
Therefore, we can use a state invariant of an LPE to strengthen all of its condition expressions.

Moreover, a state invariant of the original LPE can \emph{also} be used to restrict the behaviour of the components obtained from a cleave, as formalised in the following theorem.
Note that the set of indices is used to only strengthen the condition expressions of summands that introduce synchronisation,  because the condition expressions of independent summands cannot contain the other parameters as free variables.
Furthermore, the restriction can be applied to independent summands before the decomposition.
The theorem below states that the validity of the decomposition does not change by strengthening the components (induced by separation tuples) using state invariants.
A proof can be found in the technical report~\cite{Laveaux20}.

\begin{restatable}{theorem}{thminvariant}\label{theorem:invariant}
  Let $P(\vec{d} : \vec{D}) = \bigplus_{i \in I} \sum_{e_i : E_i} c_i \rightarrow \multiact_i \sequential P(\vec{g}_i)$ be an LPE and $(V, \Iind^V, \Isub^V, c^V, \multiact^V, h^V)$ and $(W, \Iind^W, \Isub^W, c^W, \multiact^W, h^W)$ be separation tuples as defined in Definition~\ref{def:derived}.	
	Let $\psi$ be a state invariant of $P$.	
	Given a closed expression $\vec{\initval} : \vec{D}$ such that $\interpret{[\vec{d} \gets \vec{\initval}](\psi)}$ holds, the following expression, where $C = \Isub^V \cap \Isub^W$, is a valid decomposition:
 \begin{align*}
   \hide_{\{\acttag\}} (
    	\allow_{\{\nodata{\multiact_i} \mid i \in I\} \cup \{\nodata{\multiact_i|\acttag} \mid i \in (Iind_V \cup \Iind_W)\}}( \hide_{\{\actsync^i \mid i \in I\}} ( \communication_{\{ \actsync^i_V \mid \actsync^i_W \rightarrow \actsync^i \mid i \in I\}} (P^{\psi,C}_V(\subvector{\vec{\initval}}{V}) \parallel P^{\psi,C}_W(\subvector{\vec{\initval}}{W})))
   	  )
     )
 \end{align*}

\end{restatable}

Observe that the predicate $n \leq 3$ is a state invariant of the LPE $\procmachine$ in Example~\ref{example:delayedmachine}.
Therefore, we can consider the process $\procmachine^{\psi,I}_W$ in Example~\ref{example:naivecleave} for the composition expression, which is finite.
This would yield two finite components, but the state space of $\procmachine^{\psi,I}_W$ is larger than that of $P_W$ in Example~\ref{example:refinedcleave}.

Finally, we remark that the restricted state space contains deadlock states whenever the invariant does not hold.
These deadlocks can be avoided by applying the invariant to the update expression of each parameter instead of the parameter itself without affecting the correctness.

\section{Implementation}\label{section:implementation}

While Theorems~\ref{theorem:refinedcleave} and~\ref{theorem:invariant}, and Definition~\ref{def:requirements} together provide the conditions that guarantee that a cleave yields a valid decomposition,
requirements (\REQ{3}) and (\REQ{4}) of the latter definition are difficult to ensure due to the semantic nature of these requirements.
In practice, we need to effectively approximate these correctness requirements using static analysis.

While we leave it to future work to investigate to what extent a precise and efficient static analysis is possible,
we have implemented an automated prototype translation that, given a user-supplied partitioning of the process parameters of the monolithic LPE, exploits a simple static analysis to obtain components that are guaranteed to satisfy the requirements of a cleave.
First of all the prototype identifies the independent summands in both components.
Furthermore, we decide on each clause of a conjunctive condition and action in the action expression where it belongs.
This analysis is based on the observation that whenever all free variables of an expression occur in one component then that expression should be kept in that component, and thus be removed from the other component.

\section{Case Study}\label{section:casestudy}

We have used our prototype to carry out several experiments using specifications
written in the high-level language mCRL2~\cite{GrooteM2014}, a process algebra generalising the one of Section~\ref{sec:clpe}.
To apply the decomposition technique we use the LPEs that the mCRL2 toolset~\cite{BunteGKLNVWWW19} generates as part of the pre-processing step the toolset uses before further analyses of the specifications are conducted.
We compare the results of the monolithic exploration and the exploration based on the decomposition technique.
The sources for these experiments can be obtained from the downloadable artefact~\cite{artifact}.

\subsection{Alternating Bit Protocol}

The alternating bit protocol (\abp) is a communication protocol that uses a single control bit, which is sent along with the message, to implement a reliable communication channel over two unreliable channels~\cite{GrooteM2014}.
The specification contains four processes: one for the sender, one for the receiver and two for the unreliable communication channels.

First, we choose the partitioning of the parameters such that one component ($\abp_V$) contains the parameters of the sender and one communication channel, and the other component ($\abp_W$) contains the parameters of the receiver and the other communication channel.
See Table~\ref{table:sizes} for details concerning their state spaces.
We observe that component $\abp_V$ is already larger than the original state space and that it cannot be minimised further, illustrating that traditional compositional minimisation is, in this case, not particularly useful.
However, the composition of the minimised components, listed under $\abp_V || \abp_W$, shows that it is possible to derive a (slightly) smaller state space.

\begin{table}[h]
\caption[Table]{Metrics for the alternating bit protocol.}\label{table:sizes}
\centering
\begin{tabular}{ l r r r r}
  \toprule
  Model & \multicolumn{2}{c}{original} & \multicolumn{2}{c}{minimised} \\  \cmidrule(r){2-3} \cmidrule(r){4-5} 
        & \#states & \#trans & \#states & \#trans \\
  \abp   & 182 & 230 & 48 & 58 \\ \hline
  $\abp_V$ & 204 & 512 & 204 & 512 \\ 
  $\abp_W$ & 64 & 196 & 60 & 192 \\
  $\abp_V \parallel \abp_W$ & 172 & 220 & 48 & 58 \\
  $\abp^\psi_V$ & 52 & 90 & 22 & 44 \\
  $\abp^\psi_W$ & 22 & 44 & 20 & 42 \\
  $\abp^\psi_V \parallel \abp^\psi_W$ & 172 & 220 & 48 & 58 \\
  $\abp'_V$ & 5 & 35 & 5 & 35 \\
  $\abp'_W$ & 78 & 126 & 28 & 46 \\
  $\abp'_V \parallel \abp'_W$ & 76 & 90 & 48 & 58 \\
  \bottomrule
\end{tabular}
\end{table}

The main reason for this disappointing result is because the behaviour of each process heavily depends on the state of the other processes, resulting in large components, as this information is lost in the decomposition.
We can encode such global information as a state invariant based on the \emph{control flow} parameters.
The second cleave ($\abp^\psi_V \parallel \abp^\psi_W$) for the same parameter partitioning is obtained by restricting the components using this invariant.
This does yield a useful decomposition as the state spaces of these components are both smaller than the original state space, even though their composition has again a state space that is only fractionally smaller than that of the monolithic LPE.
Finally, we have obtained a cleave into components $\abp'_V$ and $\abp'_W$ where the partitioning is not based on the original processes.
This yields a very effective cleave as shown in Table~\ref{table:sizes}.

\subsection{Practical Examples}

\begin{table}[h]
\caption[Table]{State space metrics for various practical specifications.}\label{table:hesselink}

\centering
\begin{tabular}{ l l r r r r }
  \toprule
  Model & Ref &\multicolumn{2}{c}{exploration} & \multicolumn{2}{c}{minimised} \\ \cmidrule(r){3-4} \cmidrule(r){5-6} & & \#states & \#transitions & \#states & \#transitions \\
  $\chatbox$   & \cite{RomijnS98} & 65\,536 & 2\,621\,440 & 16 & 144 \\ \hline
  $\chatbox_V$ & & 128 & 4\,352 & 128 & 3\,456  \\
  $\chatbox_W$ & & 512 & 37\,888 & 8 & 440  \\
  $\chatbox_V \parallel \chatbox_W$ & & 1\,024 & 22\,528 & 16 & 144 \\ \\
  $\hesselink$ & \cite{Hesselink98} & 914\,048 & 1\,885\,824 & 1\,740 & 3\,572 \\ \hline
  $\hesselink_V$ & & 464 & 10\,672 & 464 & 10\,672 \\
  $\hesselink_W$ & & 97\,280 & 273\,408 & 5\,760 & 16\,832 \\
  $\hesselink_V \parallel \hesselink_W$ & & 76\,416 & 157\,952 & 1\,740 & 3\,572 \\ \\
  $\WMS$ & \cite{RemenskaWVFTB12} & 155\,034\,776 & \hspace{3mm} 2\,492\,918\,760 & \hspace{3mm} 44\,526\,316 & \hspace{3mm} 698\,524\,456 \\ \hline
  $\WMS_V$ & & 212\,992 & 5\,144\,576 & 212\,992 & 2\,801\,664 \\
  $\WMS_W$ & & 1\,903\,715 & 121\,945\,196 & 414\,540 & 26\,429\,911 \\
  $\WMS_V \parallel \WMS_W$ & & 64\,635\,040 & 1\,031\,080\,812 & 44\,526\,316 & 698\,524\,456 \\
  \bottomrule
\end{tabular}
\end{table}

The $\chatbox$ specification~\cite{RomijnS98} describes a chat room facility in which four users can join, leave and send messages.
This specification is interesting because it is described as a monolithic process, which means that compositional minimisation is not applicable in the first place.
The size of the components ($\chatbox_V$ and $\chatbox_W$) before and after minimisation modulo strong bisimulation are presented to show that these are small and can be further reduced.
The composition $(\chatbox_V \parallel \chatbox_W)$ shows that indeed the decomposition technique can be used quite successfully, because the result under exploration is much smaller than the original state space.
Finally, we have also listed the size of the minimised original state space (which is equal to the minimised composition) as it indicates the best possible result.
The $\hesselink$ specification~\cite{Hesselink98} describes a wait-free handshake register and the $\WMS$ specification a workload management system~\cite{RemenskaWVFTB12}, used at CERN.
For the latter two experiments we found that partitioning the parameters into a set of so-called \emph{control flow parameters} and remaining parameters yields the best results.

We also consider the total execution time and maximum amount of memory required to obtain the original state space using exploration and the state space obtained using the  decomposition technique, for which the results can be found in Table~\ref{table:metrics}.
The execution times in \textbf{s}econds or \textbf{h}ours required to obtain the state space under `exploration' in Table~\ref{table:hesselink}, excluding the final minimisation step of the original or composition state space which are only shown for reference.
The cost of the static analysis of the cleave itself was in the range of several milliseconds.

\begin{table}[h]
\caption[Table]{Execution times and maximum memory usage measurements.}\label{table:metrics}
\centering
\begin{tabular}{ l r r r r }
  \toprule
Model & \multicolumn{2}{c}{monolithic} & \multicolumn{2}{c}{decomposition} \\
 \cmidrule(r){2-3} \cmidrule(r){4-5} 
 & time & memory & time & memory \\
 $\chatbox$ & \hspace{3mm} 4.76s & \hspace{3mm} 21.9MB & \hspace{3mm} 0.2s & \hspace{3mm} 15.7MB \\
 $\hesselink$ & 7.94s & 99.7MB & 1.56s & 47.7MB \\
 $\WMS$ & 2.4h & 15.1GB & 0.8h & 11.8GB \\
\bottomrule
\end{tabular}
\end{table}

\subsection{Connect Four}

The $\connectfour$ specification models the behaviour of a game played by two players on a board with seven columns and four rows.
For this specification we can show that the decomposition can be applied recursively to obtain more than two components.
Using the decomposition procedure we first obtain a monolithic process for the left-most column and one process for the six remaining columns.
Next, we apply the decomposition to the process for the six remaining columns until we have one monolithic process for every column.
In Table~\ref{table:connectfour} we can see the state space of the process for only column seven, which is similar to the size of the other components.
We then compose columns six and seven, which is the state space listed under Columns $6\cdots7$ and the composition of column five, six and seven is listed under Columns $5\cdots7$, \emph{etcetera}.
Repeating this process until we have composed all the columns shows that we can obtain a state space that is roughly a quarter in size compared to the original state space in the number of states and transitions.
Both the monolithic exploration and the decomposition take about 30 hours.
However, the monolithic exploration required about 500GB memory whereas the decomposition requires about 234GB of main memory.
Furthermore, the state space is also immediately smaller.

\begin{table}[H]
\caption[Table]{State space metrics for the connect four specification.}\label{table:connectfour}
\centering
\begin{tabular}{l r r r r}
  \toprule
  Model & \multicolumn{2}{c}{original} & \multicolumn{2}{c}{minimised} \\  \cmidrule(r){2-3} \cmidrule(r){4-5} 
        & \#states & \#trans & \#states & \#trans \\
        
  Connect Four & 4\,571\,392\,011 & 17\,968\,443\,566 & 418\,390\,653 & 2\,079\,589\,075 \\ \hline
  Column 7   & 31  & 1\,664 & 31 & 161 \\
  Columns $6\cdots7$ & 961 & 7\,853 & 961 & 7\,751  \\
  Columns $5\cdots7$ & 29\,791 & 298\,579 & 22\,821 & 230\,003 \\
  Columns $4\cdots7$ & 707\,451 & 8\,179\,465 & 336\,537 & 4\,138\,521 \\
  Columns $3\cdots7$ & 11\,362\,647 & 112\,362\,633 &  6\,112\,522 & 60\,088\,448 \\
  Columns $2\cdots7$ & 189\,489\,112 & 1\,824\,751\,492 & 92\,251\,708 & 908\,605\,682 \\
  Columns $1\cdots7$ & 1\,049\,255\,356 & 4\,889\,577\,305 & 418\,390\,653 & 2\,079\,589\,075 \\  
  \bottomrule
\end{tabular}
\end{table}

\section{Conclusion}\label{section:conclusion}

We have presented a decomposition technique, referred to as cleave, that can be applied to any monolithic process with the structure of an LPE and have shown that the result is always a valid decomposition.
Furthermore, we have shown that state invariants can be used to improve the effectiveness of the decomposition.
We consider defining a static analysis to automatically derive the parameter partitioning for the practical application of this technique as future work.
Furthermore, the cleave is currently not well-suited for applying the typically more useful abstraction based on (divergence-preserving) branching bisimulation minimisation~\cite{GlabbeekLT09}.
The reason for this is that $\tau$-actions might be extended with synchronisation actions and tags.
As a result they become visible, effectively reducing branching bisimilarity to strong bisimilarity.

\section*{Acknowledgement}
This work is part of the TOP Grants research programme with project number 612.001.751 (AVVA), which is (partly) financed by the Dutch Research Council (NWO).
We also would like to thank the anonymous reviewers for their effort and constructive feedback.

\bibliographystyle{eptcs}
\bibliography{bibliography}

\end{document}